\documentclass[iop,apj]{emulateapj}
\usepackage{natbib} 
\usepackage{color} 
\usepackage{multirow}
\usepackage{rotating}
\usepackage{framed} 
\usepackage{booktabs}
\usepackage{hyperref} 
\usepackage{xspace}
\usepackage{amsmath}
\usepackage{paralist}

\providecommand{\supportfrom}[1]{#1}
\newcommand{\BE}{\begin{equation} }
\newcommand{\EE}{\end{equation} }
\def\ba#1\ea{\begin{align}#1\end{align}}
\newcommand{\h}{\ensuremath{{h}}\xspace}

\newcommand{\rmin}{\ensuremath{{r_-}}\xspace}
\newcommand{\rmax}{\ensuremath{{r_+}}\xspace}
\newcommand{\rminmax}{\ensuremath{{r_\pm}}\xspace}

\newcommand{\precnone}{\ensuremath{{\Delta \phi}} \xspace}
\newcommand{\precnewt}{\ensuremath{{\Delta \phi_{\rm Newt.}}} \xspace}
\newcommand{\precgr}{\ensuremath{{\Delta \phi_{\rm GR}}} \xspace}
\newcommand{\precerr}{\ensuremath{{\delta \phi}} \xspace}

\newcommand{\potf}{\ensuremath{{U( r )}} \xspace}
\newcommand{\espace}{\,}
\newcommand{\pw}{Paczy{\'n}ski-Wiita\xspace}
\newcommand{\pwr}{\citet{Paczynsky:1980}\xspace}
\newcommand{\OO}{\ensuremath{\mathcal{O}}}

\newcommand{\myemail}{wegg@tapir.caltech.edu}
\shorttitle{Potentials for Parabolic Orbits}
\shortauthors{Wegg}
\begin{document}
\title{Pseudo-Newtonian Potentials for Nearly Parabolic Orbits}
\author{Christopher\ Wegg\altaffilmark{1}}
\affil{Theoretical Astrophysics, California Institute of
Technology, MC 350-17, 1200 East California Boulevard,
Pasadena, CA 91125}
\altaffiltext{1}{\href{mailto:\myemail}{\myemail}}  

\begin{abstract}

We describe a pseudo-Newtonian potential which, to within $1\%$ error at all angular momenta, reproduces the precession due to general relativity of particles whose specific orbital energy is small compared to $c^2$ in the Schwarzschild metric. For bound orbits the constraint of low energy is equivalent to requiring the apoapsis of a particle to be large compared to the Schwarzschild radius. Such low energy orbits are ubiquitous close to supermassive black holes in galactic nuclei, but the potential is relevant in any context containing particles on low energy orbits.
Like the more complex post-Newtonian expressions, the potential correctly reproduces the precession in the far-field, but also correctly reproduces the position and magnitude of the logarithmic divergence in precession for low angular momentum orbits. An additional advantage lies in its simplicity, both in computation and implementation. 
We also provide two simpler, but less accurate potentials, for cases where orbits always remain at large angular momenta, or when the extra accuracy is not needed. In all of the presented cases the accuracy in precession in low energy orbits exceeds that of the well known potential of \pwr, which  has $\sim30\%$ error in the precession at all angular momenta. 
\end{abstract}

\section{Introduction} 
\label{introduction}

Pseudo-Newtonian potentials that modify the Newtonian gravitational potential have a long history of use in astrophysics. While general relativity is now well understood in the astrophysics community, pseudo-Newtonian potentials are still useful in approximating relativistic effects in simpler and faster Newtonian simulations. The potential of \pwr is often used in the study of accretion onto relativistic objects. In this regime the \pw potential often gives results close to those using full GR since it correctly reproduces the location of the inner most stable circular orbit (ISCO) and the marginally bound orbit as well as being a good approximation to the binding energy at the ISCO \citep[for a review see][]{Abramowicz:2009}.

Here we propose a series of Newtonian potentials with a different aim: to correctly reproduce the precession produced by general relativity in the Schwarzschild metric for test particles whose apoapsis lies far from the hole, i.e., in the non-relativistic region. The \pw potential has been used in this context multiple times \citep[e.g.][]{Chen:11}, despite its key properties of closely reproducing the location and energy of the ISCO being unimportant in this regime. Instead, we propose alternative potentials that are more accurate and physically relevant for these orbits. We have used them to simulate galactic dynamics around supermassive black holes (SMBHs) in \citet{bode:11pre}. The primary concern in that context was to ensure that stars passing close to the black hole exited along the correct trajectories. These potentials are likely to be useful in other contexts, motivating the brief presentation here. Throughout we use geometrized units where $G=c=1$.

\begin{figure}
\centering
\includegraphics[width=0.99\columnwidth]{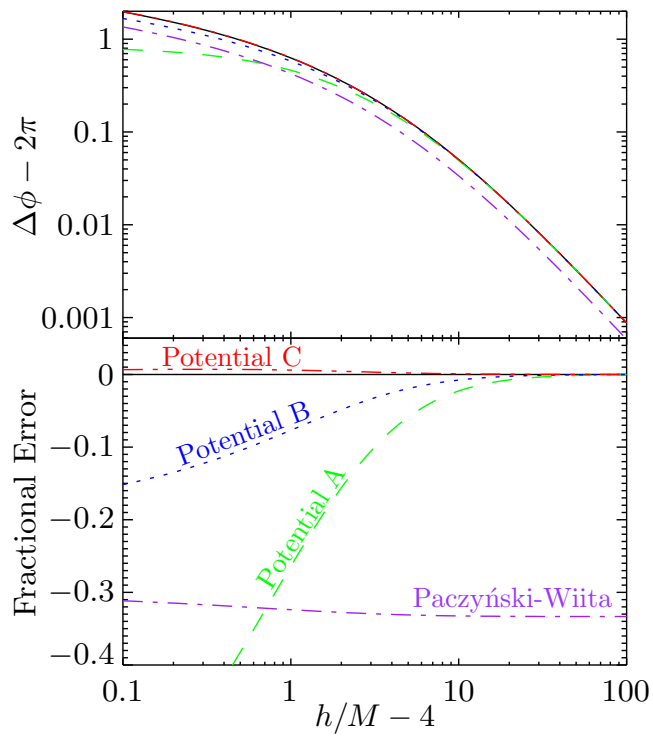}
\caption{Comparison of the precession per orbit produced by the proposed potentials with the GR expression for parabolic orbits as a function of specific angular momentum, $h$. The labeled potentials are described in Table \ref{tab:potcoeffs}. The GR precession is shown by a solid black line (in the upper panel the precession produced by potential C lies almost on top of the GR expression).  In the lower panel we plot the fractional error relative to the relativistic precession, defined to be $(\precnone - \precgr)/(\precgr - 2\pi)$. }
\label{fig:prec}
\end{figure}

\begin{table*}
\begin{center}
\caption{Coefficients for the potentials $\potf = -{\alpha M}/{r} -{(1-\alpha)M}/{(r-R_x)} -{M R_y}/{r^2}$ (equation \ref{eqn:pngen}) described in this work.}
    \begin{tabular}{  c  c  c  c c c c}
    \hline
Potential & $\alpha$ & $R_x/M$ & $R_y/M$ & Precession error: $r\gg M$ & Diverges at $h = 4M$&Maximum precession error
\\ \hline
A & 1 & --- & 3 & $ 0 $& No&$100\%$\tablenotemark{a} \\
B & 0 & 5/3 & 4/3 &$ 0 $& Yes& $<30\%$\\
C & $\frac{-4}{3}\left(2+\sqrt{6}\right)$ & $\left(4\sqrt{6}-9\right)$ & $\frac{-4}{3}\left(2\sqrt{6}-3\right)$ &$0$&Yes& $<1\%$ \\
\pw & 0 & 2 & 0 &$33\%$&Yes& $33\%$ \\
\hline
\label{tab:potcoeffs}
\end{tabular}
\footnotetext[1]{Does not diverge at $h=4M$}
\end{center}
\end{table*}

\section{Summary of Proposed Potentials}
We present three new pseudo-Newtonian potentials in this paper. All of these potentials, and the potential of \pwr, can be written in the form
\BE
\potf = -\frac{\alpha M}{r} -\frac{(1-\alpha)M}{r-R_x} -\frac{M R_y}{r^2} \espace .
\label{eqn:pngen}
\EE
where the values of the coefficients $\alpha$, $R_x$ and $R_y$ for the potentials are summarized in Table \ref{tab:potcoeffs}. We choose potentials of this form since the presence of the $1/r^2$ term allows the far field behavior to reproduced, while the $1/(r-R_x)$ term allows reproduction of the divergent behavior as the specific angular momentum approaches $4M$. The resultant precession per orbit is compared to the GR value in Figure \ref{fig:prec}. In what follows we justify these choices.

\section{Approach To Calculating Proposed Potentials}
\subsection{Precession Due to General Relativity}

In general relativity the change in azimuthal angle of a test particle between two consecutive apoapsides on a geodesic in the Schwarzschild metric is given by \citep[e.g. Equation 25.42 of][]{Misner:73} 
\BE 
\precgr = 2 \int\limits_\rmin^\rmax \left[ \frac{(E+1)^2}{h^2} - \left(\frac{1}{h^2} + \frac{1}{r^2}\right) \left(1-\frac{2 M}{r} \right) \right]^{-1/2} \frac{dr}{r^2} 
\label{eqn:precgen} \espace , 
\EE
where $E$ is the specific energy of the particle without rest mass energy (i.e. $E\equiv -p_0/\mu - 1$ where $\mu$ is the particles mass), $h$ is the specific angular momentum (i.e. $h=p_\phi/\mu$), and \rminmax are the radii of periapsis ($-$) and apoapsis ($+$) given by the two largest roots of the equation
\BE
(E+1)^2 -  \left( 1 - \frac{2 M}{\rminmax} \right) \left( 1 + \frac{h^2}{\rminmax^2} \right) = 0
\label{eqn:rminmaxgen} \espace .
\EE
For our `nearly parabolic orbits' ($E\ll 1$) the precession due to relativity is a function only of the angular momentum. 
Unless otherwise noted in this paper we work in the limit that $E=0$.

\subsection{Precession due to Newtonian Central Potential}	

By comparison, in classical mechanics the change in azimuthal angle for a test particle of any energy between two consecutive apoapsides in a central potential, $U( r )$, is given by \citep[e.g.][]{Landau:Mechanics}
\BE
\precnewt= 2 \int\limits_\rmin^\rmax \left[ \frac{ E - U( r ) }{h^2/2} - \frac{1}{r^2} \right]^{-1/2} \frac{dr}{r^2}
\label{eqn:precnewt} \espace , 
\EE
where in this case \rminmax are given by
\BE
2 ( E - U( \rminmax ) ) - \frac{h^2}{\rminmax^2} = 0 
\label{eqn:rminmaxnewt} \espace .
\EE

\subsection{Requirements of Proposed Pseudo-Newtonian Potentials}

In principle it is possible to define a pseudo-Newtonian potential, $U( r )$, such that the precession angles given by equations \ref{eqn:precgen} and \ref{eqn:precnewt} are equal in the limit $E\ll 1$, i.e.
\BE
\precgr ( h ) = \precnewt ( h ) \label{eqn:desired} \espace .
\EE
This potential would have the property desired: on returning to large radii, test particles would have precessed through the correct angle, and be traveling along the correct path with only a time error. However, we also desire a simple potential for efficient calculation and so instead we seek potentials that minimize the precession error, $\delta \phi$, defined through
\BE
\precerr ( h ) = \precgr ( h ) - \precnewt ( h ) \label{eqn:errordef} \espace .
\EE

We propose three potentials that, in order of complexity, minimize $\precerr$:
\begin{inparaenum}[\itshape A\upshape)]
\item  in the far field (large $h$) ;
\item  in the far field {\it and} whose precession diverges logarithmically in the same location as GR ($h\to4 M$); and
\item  in the far field, and whose precession diverges logarithmically as $h\to4 M$ {\it with the correct magnitude}.
\end{inparaenum}

\section{Proposed Pseudo-Newtonian Potentials}
 

\subsection{Potential A: Matching The Far Field Precession}

In this section we consider the behavior of orbits with $h\gg4M$, but we do not require $E=0$, only that $E\ll 1$. In this case,  inspection of Equation \ref{eqn:rminmaxgen}  
shows $\rminmax \gg M$ and  the entire orbit lies in the far field.

The change in angle in the far field in GR can be calculated from equation \ref{eqn:precgen} and is well known to be \citep[e.g.][]{Weinberg:72}
\BE
\precgr (\h) = 2\pi +  \frac{6 \pi M^2}{\h^2} \quad {\rm for} \quad \h\gg 4M \espace .
\label{eqn:grfar}
\EE
Note that all that is required is a sufficiently distant periapse. It is not required that the orbit have $E=0$.

In the far field we require potentials have the form
\BE
\potf = - \frac{M}{r} - \frac{M R_y}{r^2} + \OO(r^{-3}) \espace .
\label{eq:farexpand}
\EE
Neglecting the terms $\OO(r^{-3})$ and higher, after some algebra, the precession calculated from equation \ref{eqn:precnewt} is 
\BE
\precnewt ( \h ) = 2 \int\limits_\rmin^\rmax \left[ \frac{2 r_h a}{(r - \rmin)(\rmax-r)} \right]^{1/2} \frac{dr}{r}
\label{eqn:pnfarint} \espace ,
\EE
where $r_h\equiv{h^2}/{2 M}$, $a\equiv{-M}/{2E}$, and
\BE
\rminmax = a\left[1\pm\sqrt{1-\frac{2}{a}(R_y-r_h)}\right] \espace .
\EE 
Contour integration gives the integral
\ba
\precnewt ( \h ) 
&= 2\pi \left(1-\frac{R_y - a}{r_h}\right)^{-1/2} \nonumber \\
&= 2\pi + \frac{2\pi M R_y}{h^2} +\OO(a/r_h) \espace .
\ea
Provided that $h \gg M$ and $E\ll 1$, then the final term can be dropped and matching the far field precession given by equation \ref{eqn:grfar} requires $R_y=3M$. 

When concerned with the far field precession we therefore propose the pseudo-Newtonian potential
\BE
\potf = -\frac{M}{r}\left( 1 + \frac{3 M}{r} \right) 
\label{eqn:pnfar} 
\EE
which requires only modest additional computation over the Newtonian expression.
Note that this pseudo-Newtonian potential gives the correct precession for {\it all} orbits (parabolic, eccentric, or hyperbolic), provided that the periapse lies in the far field. The resultant precession, labeled as potential A, is plotted in Figure \ref{fig:prec}. 

Since this potential does not reproduce the divergence as $h\to 4M$ this potential performs worse than the \pw potential, even for parabolic orbits, when $h\lesssim 4.8M$ which corresponds to periapsis separation of $r \lesssim 8.5M$. For reference $h=4M$ corresponds to periapsis separation $r=5M$ for parabolic orbits in this potential.

\subsection{Potential B: Logarithmic Divergence As $\h\to 4M$}

In this section we construct a potential that reproduces the logarithmic divergence of the general relativistic precession as $\h\to4M$.

First consider a potential of the form
\BE
\potf = -\frac{M}{r-R_x} \espace .
\EE
The precession in this potential for $E=0$ is given by
\BE
\precnewt ( h ) = 2 \int\limits_\rmin^\infty \left[ \frac{r^2 }{r_h(r-R_x)} - 1 \right]^{-1/2} \frac{dr}{r}
\espace . 
\EE
The roots of the quadratic form in the integrand can be written as
\BE
r_{p,q} = \frac{r_h}{2}\left( 1 \pm \sqrt{ 1 - \frac{4R_x}{r_h}} \right) \espace .
\EE
so that
\BE
\precnewt ( h ) = 2 \int\limits_{r_p}^\infty \left[ \frac{r_h(r-R_x)}{(r - r_p)(r - r_q)} \right]^{1/2} \frac{dr}{r} \espace .
\EE
As $r_h\to 4 R_x$, then $r_p\to r_q$  and the integral diverges logarithmically. $r_h \to 4 R_x$ corresponds to $h\to\sqrt{8 M R_x}$ and the leading order behavior of the integral is
\BE
\lim_{r_h \to 4 R_x^+} \precnewt ( h ) = -\log \left(h - \sqrt{8 M R_x}\right) \espace .\label{eqn:newtPWdivrate}
\EE

A similar calculation using the GR expression gives a logarithmic divergence as $h\to4M$ and the corresponding expression is 
\BE
\lim_{h \to 4 M^+} \precgr ( h ) = - \sqrt{2} \log \left(h - 4 M\right) \espace .
\label{eqn:grdivrate}
\EE

For $\precnewt$ to diverge at $h=4M$, we must have $R_x=2M$. This is exactly the potential of \pwr, which therefore diverges at the correct angular momentum. However comparing equations \ref{eqn:newtPWdivrate} and \ref{eqn:grdivrate}, the potential of \pwr has the incorrect magnitude (by a factor of $\sqrt{2}$) as the angular momentum approaches $4M$, and as noted previously, has incorrect precession in the far field.

To correct the far field behavior consider the potential
\BE
\potf = -\frac{M}{r-R_x} -\frac{M R_y}{r^2} \espace .
\EE
The calculation of the precession proceeds in the same manner, but with the roots now given by
\BE
r_{p,q} = \frac{r_h - R_y}{2}\left( 1 \pm \sqrt{ 1 - \frac{4R_x}{r_h - R_y}} \right) \espace .
\EE	
Again, as the roots coincide the integral diverges logarithmically. Requiring that the divergence occurs as $h \to 4M$ and that the potential has the correct far field limit (i.e. far from the hole the expansion is given by equation \ref{eq:farexpand} with $R_y=3M$) gives $R_x={5M}/{3}$ and $R_y={4M}/{3}$. Our proposed potential which has the correct precession in the far field and which also logarithmically diverges as $h \to 4M$ is therefore
\BE
\potf = -\frac{M}{r} \left(\frac{1}{1-{5M}/{3r}}  + \frac{4M}{3 r}\right) 
\label{eqn:pndiv} \espace .
\EE
The resultant precession, labelled as potential B, is shown in Figure \ref{fig:prec}. 
This is the potential used in \cite{bode:11pre}. For reference $h=4M$ corresponds to periapsis separation $r=10M/3$ for parabolic orbits in this potential.

\subsection{Potential C: Correct Rate Of Logarithmic Divergence As $\h\to 4M$}
\label{sec:divrate}

The potential proposed in equation \ref{eqn:pndiv},  has the correct far field behavior, and diverges logarithmically at the correct angular momentum. However, the rate of that divergence is incorrect: The behavior of the integral as $h \to 4 M$ is
\BE
\lim_{h \to 4 M^+} \precnewt ( h ) = - \sqrt{\frac{6}{5}} \log \left(h - 4 M\right) \espace ,
\EE
which does not match the GR expression in equation \ref{eqn:grdivrate}.

An additional term in the potential allows this to rectified. Using a potential of the form
\BE
\potf = -\frac{\alpha M}{r} -\frac{(1-\alpha)M}{r-R_x} -\frac{M R_y}{r^2} \espace ,
\label{eqn:pngood}
\EE
enables us to match the three constraints for the three coefficients $\alpha$, $R_x$ and $R_y$. The constraints on the coefficients are that:
\begin{inparaenum}[\itshape 1\upshape)]
\item in the far field the precession approaches equation \ref{eqn:grfar} (i.e. equation \ref{eq:farexpand} with $R_y=3M$), 
\item the integral diverges logarithmically as $h\to 4M$, and
\item the rate of divergence as $h\to 4M$ is given by equation \ref{eqn:grdivrate}.
\end{inparaenum}
The values of $\alpha$, $R_x$ and $R_y$ satisfying these constraints are
\ba
\alpha &=	\frac{-4}{3}\left(2+\sqrt{6}\right) 	\espace , \nonumber \\
R_x &=		\left(4\sqrt{6}-9\right)M 				\espace , \\
R_y &=		\frac{-4}{3}\left(2\sqrt{6}-3\right)M	\espace . \nonumber
\ea
The precession produced by this ``potential C'' is compared to the GR expression in Figure \ref{fig:prec}. For reference $h=4M$ corresponds to periapsis separation of $r=2(\sqrt{6}-1)M$ for parabolic orbits in this potential.

This potential produces precession which agrees with GR to within $1\%$ for  {\it all} orbits where $E \ll 1$ i.e. whose specific orbital energy is small compared to $c^2$ in the Schwarzschild metric. For bound orbits this corresponds to requiring apoapsis be large compared to the Schwarzschild radius ($r_+\gg M$).  

\section{Conclusions}

We have proposed three pseudo-Newtonian potentials appropriate for `nearly parabolic orbits' (orbital energy, $E \ll 1$) around a Schwarzschild black hole. These nearly parabolic orbits correspond to orbits with large apoapsis compared to the Schwarzschild radius of a central black hole, or mildly hyperbolic orbits. 

For bodies which pass close to the black hole, these potentials accurately reproduce the changes in the ``Newtonian'' parts of the trajectories far from the black hole, differing from the exact GR expression only by a time error. 
In the far field the time error as a fraction of the orbital period is of order $\delta P/P = \OO(E) \ll 1$ but diverges as $h\to4M$ like $\delta P/P = \OO(E^{3/2} \log(h-4M))$. Therefore, for $E \ll 1$, the fractional period error is small outside of an exponentially narrow region in $h$ close to $4M$.

The potentials reproduce general relativistic precession with varying degrees of accuracy and simplicity. Namely, these potentials produce accurate relativistic precession:
\begin{inparaenum}[\itshape A\upshape)]
\item[(Potential A)]  in the far field (equation \ref{eqn:pnfar});
\item[(Potential B)]  in the far field {\it and} with the logarithmic divergence as $h\to4 M$ (equation \ref{eqn:pndiv}); and
\item[(Potential C)]  in the far field {\it  and with the correct magnitude} of logarithmic divergence as $h\to4 M$ (equation \ref{eqn:pngood}).
\end{inparaenum}

The potentials described do not include the effects of spin, or gravitational radiation, which can be astrophysically important for orbits passing close to the hole. Neither of these effects can be described by a pseudo-Newtonian potential without the presence of undesirable inseparable terms including both $r$ and $v$, and so were not considered in this work. For objects whose mass ratio with the black hole is sufficiently close to the test particle limit, the effects of gravitational radiation can be included by subtracting energy and angular momentum at periapsis, for example using the results of \cite{Gair:06}. 

Close to the black hole these potentials should be interpreted with care since although they diverge at the correct angular momentum, the $r$ at which this occurs does not correspond to the Schwarzschild radial coordinate. 

\acknowledgments
We gratefully acknowledge many useful discussions with Nate Bode, Laura Book, Kevin Setter, and Sterl Phinney.

Support for this work was provided by NASA BEFS grant \supportfrom{NNX-07AH06G}.

\bibliographystyle{apjads}

\end{document}